\documentstyle[prl,aps,epsfig,amssymb,multicol]{revtex}
\begin{document}
\draft \title{Canonical form of Hamiltonian matrices} 
\author{ A. P. Zuker, L. Waha Ndeuna, F.
  Nowacki, and E. Caurier,} 
\address{ IRES, B\^at27,
  IN2P3-CNRS/Universit\'e Louis Pasteur BP 28, F-67037 Strasbourg
  Cedex 2, France} 
\date{\today} 
\maketitle
\begin{abstract}
  On the basis of shell model simulations, it is conjectured that the
  Lanczos construction at fixed quantum numbers defines---within
  fluctuations and behaviour very near the origin---smooth canonical
  matrices whose forms depend on the rank of the Hamiltonian,
  dimensionality of the vector space, and second and third
  moments. A framework emerges that amounts to a general Anderson
  model capable of dealing with ground state properties and strength
  functions. The smooth forms imply binomial level densities. A
  simplified approach to canonical thermodynamics is proposed.
\end{abstract}
\begin{multicols}{2}

\begin{figure}[htb]
 \begin{center}
 \leavevmode
  \epsfig{file=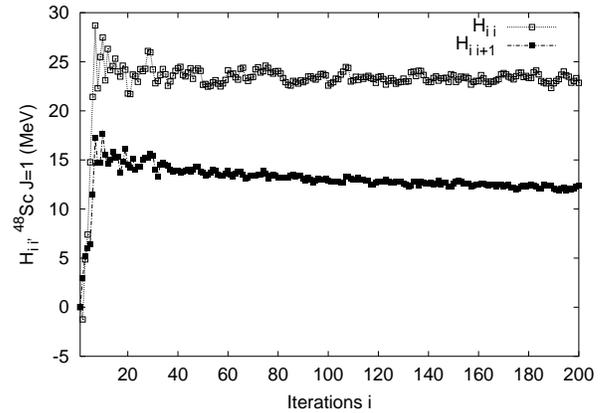, width=8.cm }
 \caption{Exact matrix elements for $^{48}$Sc
   ($J=1$)}
 \label{fig:sc48}
 \end{center}
\end{figure}

Since many physical problems can be thought as matrix
diagonalizations, much attention has been devoted to the general
properties of matrices, starting with Wigner's Gaussian Orthogonal
Ensemble (GOE) which has provided invaluable insight into fluctuation
properties~\cite{por65,bro81}. However, physical systems are usually
associated to Hamiltonians of rank ($r$) 1+2 (i. e., 1+2-body) while
for GOE the rank is the dimensionality of the space.  Two Body Random
Ensemble (TBRE) or Embedded GOE (EGOE) are the suggested denominations
for limited ranks. We shall prefer to speak of Hamiltonian matrices,
either random or not. Exact results and even numerical ones are
hampered by the difficulty of enforcing the rank conditions, which are
hard to simulate through special band or sparse matrices, and it would
be desirable to have a canonical form that ensures rank 1+2. It is our
aim to show that the Lanczos tridiagonal construction can provide it.
(The impatient reader may proceed to the discussion after
Fig.~\ref{fig:sc48}.)

The algorithm is standard~\cite{wil65}:
 
A starting vector (the pivot) $|0\rangle$ is chosen. The vector
$|a_0\rangle ={\cal H} |0\rangle$ has necessarily the form
$|a_0\rangle=H_{00}|0\rangle +|1'\rangle>$, with $\langle 0|1'\rangle
=0$. Calculate $\langle 0|a_0\rangle=H_{00}$. Normalize $|1'\rangle$,
to find $H_{01}=\langle 1'|1'\rangle ^{1/2}$. Iterate until state
$|k\rangle$ has been found. The vector $|a_k\rangle ={\cal H}
|k\rangle$ has necessarily the form
\begin{equation}
  \label{lanc}
  |a_k\rangle= H_{k\, k-1} |k-1\rangle+ H_{k\, k} |k\rangle   
                        + |(k+1)'\rangle, 
\end{equation}
Calculate $\langle
  k|a_k\rangle=H_{k\, k}$. Extract and normalize $|(k+1)'\rangle$, to find
$H_{k\, k+1} = \langle (k+1)'|(k+1)'\rangle  ^{1/2}$.

The choice of pivot is arbitrary and it can be adapted to special
problems. One of the most interesting is the calculation of strength
functions: Given a transition operator ${\cal T}$, act with it on a
target state $|t\rangle$ to define a pivot $|0\rangle={\cal
  T}|t\rangle$ that exhausts the sum rule for ${\cal T}$. The
tridiagonal matrix elements are then linear combinations of the
moments of the strength distribution $\langle 0|{\cal H}^K|0\rangle$
(expand $|0\rangle$ in eigenstates to check the statement).  Therefore
the eigensolutions of the $I\times I$ matrix define an approximate
strength function $S_I(E)=\sum_{i=1,I}\, \delta(E-E_i)\, \langle
i|{\cal T}|0\rangle^2$, whose first $2I-1$ moments are the exact ones
(state $|I\rangle$ is obtained by orthogonalizing ${\cal
  H}^{I-1}|0\rangle$ to the preceeding states. Hence $H_{I\,
  I}=\langle 0|{\cal H}^{2I-1}|0\rangle$ + information from previous
iterations) The eigenstates act as ``doorways'' whose strength will be
split until they become exact solutions for $I$ large enough. To fix
ideas consider in Fig.~\ref{fig:sc48} the first 200 Lanczos matrix
elements calculated to reproduce the Gamow Teller strength (${\cal
  T}=\sigma \tau$) obtained through the reaction
$^{48}$Ca(p,n)$^{48}$Sc. In Ref~\cite{cpz95} 700 iterations proved
necessary to ensure the exact $S(E)$ in the $pf$ shell (the dimension
of the $JT=13$ space is $d=8590$). The question at the origin of the
present work is: do we truly need 700 exact eigensolutions to obtain a
good $S(E)$? Probably not, and the idea is to calculate exactly a few
matrix elements, and replace the rest by an adequate approximation.
The very rapid variation of the first matrix elements in
Fig.~\ref{fig:sc48} certainly depends on the pivot. It is followed by
fluctuations around smooth patterns: a constant for the diagonals, and
a line $b$ for the off diagonals. Therefore, a good approximation to
the exact problem comes in three steps. First determine the smooth
patterns, then do the few Lanczos iterations necessary to reach them
and finally replace the rest of the matrix elements by fluctuations
mounted on the smooth patterns. 

Next we concentrate on the first step.

The quantities that turn out to be necessary are the dimensionality,
which we write as $d=2^N$, the centroid $E_c=d^{-1}\text{tr}(H)$, the
variance $\sigma^2$ and skewness coefficient $\gamma_1$:
\begin{eqnarray}
\label{eq:mom01}
d&=&2^N \Longrightarrow N=\ln d/\ln 2,\qquad E_c=d^{-1}\text{tr}(H)\\ 
  \sigma^2&=&d^{-1}\text{tr}(H-E_c)^2,\;
  \gamma_1=(d\sigma^3)^{-1}\text{tr}(H-E_c)^3.\label{eq:mom23}  
\end{eqnarray}

An inverse binomial function is also needed: 

$x=\text{ nib}(b,N), \; 0<x<1/2$,
the inverse of
 \begin{equation}
   \label{eq:bin}
b=\text{bin}(x)=e^{\frac{1-N}{2}\left[(1+2x)\ln(1+2x)+(1-2x)\ln(1-2x)\right]},
 \end{equation}
the Stirling approximation of
$$b=\text{
   bin}(x,N)=2^{-N}{N\choose N(1/2-x)},$$ a binomial shifted and
normalized so as to be maximum and unity at the origin. 

The ``data'' from which we draw conclusions are exact
diagonalizations (at fixed angular momentum $J$) for $^{48}$Ca, a
typical closed shell nucleus~\cite{cau94}, and for the configurations
$(f_{7/2}p_{3/2})^4_{\pi}(g_{9/2}d_{5/2})^4_{\nu}$ ( $(fpgd)^8$ for
short) that exhibit rotational behaviour with backbend~\cite{zuk95}.

\begin{figure}[htb]
 \begin{center}
 \leavevmode
  \epsfig{file=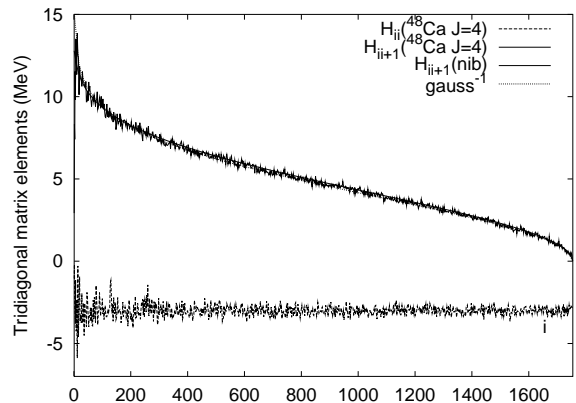, width=8.cm }
 \caption{Exact matrix elements for $^{48}$Ca
   ($J=4$), the inverse binomial nib$\equiv H_{ii+1}$
   (Eqs.~(\ref{eq:Hii'})) , and  Gauss$^{-1}\equiv
   H_{ii+1}^G$ (Eq.(~\ref{eq:G})). Parameters: $d=1755$, $N=10.78$, $\sigma=7.85$
   MeV, $\gamma_1=-.01$. See text}
 \label{fig:nibs}
 \end{center}
 \end{figure}

\begin{figure}[htb]
\begin{center}
   \epsfig{file=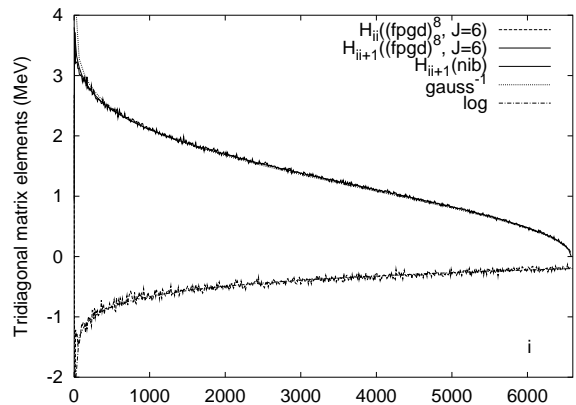, width=8.cm } 
    \caption{Exact matrix elements (only one every ten are plotted) for
      $(fpgd)^8$ ($J=6$) compared with log$\equiv H_{ii}$
      (arbitrarily shifted), and nib$\equiv H_{ii+1}$ from
      Eqs.~(\ref{eq:Hii'}), and Gauss$^{-1}\equiv H_{ii+1}^G$
      (Eq.~(\ref{eq:G})) . Parameters: $d=6579$, $N=12.68$, $\sigma'=2.14$ MeV,
      $\sigma=2.18$, $\gamma'_1=-0.246$, $\gamma_1=-0.296$. See text}
   \label{fig:niba}
 \end{center}
 \end{figure}

The matrix elements in Figs.~\ref{fig:nibs} and~\ref{fig:niba}
correspond to maximum $d$ for each family. They are totally
representative of the other members. Only the general
(pivot-independent) trend is visible.

 To within fluctuations the diagonals are constant or logarithmic, and
 the off-diagonals have an inverse binomial form.

The matrix elements that reproduce---perfectly within
fluctuations---the exact ones are
 \begin{equation}
  \label{eq:Hii'}
H_{i\,i}=-\gamma'_1(\ln\frac{i}{d}+1), \quad H_{i\,
  i+1}=\sigma'\sqrt{N}\, \text{nib}(\frac{i}{d},N).   
\end{equation}
The subtraction in the ln term ensures that $H_{i\, i}$ is traceless.
$\sigma'$ and $\gamma'_1$ are
 determined so as to yield  $\sigma$ and $\gamma_1$  through 
\begin{eqnarray}
  \label{eq:m2}
\sigma^2&=&\langle{\cal H}^2\rangle=d^{-1}\sum H_{ii+1}^2 + H_{ii-1}^2
+ H_{ii}^2,\\ 
  \label{eq:m3}
\sigma^{3} \gamma_1&=&\langle{\cal H}^3\rangle\simeq d^{-1} \sum H_{ii}(3
 H_{ii+1}^2 + 3 H_{ii-1}^2 + H_{ii}^2).  
\end{eqnarray}
 For the symmetric case ($\gamma_1\approx 0$), the solution is simply
$\sigma'\approx \sigma$.  For the asymmetric case, Eqs.~(\ref{eq:m2}
and \ref{eq:m3}) have to be solved. All parameters are given in the
captions to Figs.~\ref{fig:nibs} and~\ref{fig:niba}

An approximation to the nib function is Gauss$^{-1}=x=\sqrt{-(2N)^{-1}
  \ln{\, g}}\, $, the inverse of $g=\exp(-2Nx^2)$ obtained by
expanding $b$ in Eq.~(\ref{eq:bin}) around $x=0$. Therefore
$H_{i\, i+1}$ in Eq.~(\ref{eq:Hii'})is approximated by
\begin{equation}
\label{eq:G}
H_{i\, i+1}^G=\sqrt{-(\sigma')^2/2\ln(i/d)},
\end{equation}
which is seen in Figs.~\ref{fig:nibs} and ~\ref{fig:niba} to be fairly
close to nib except for the severe overshooting at the origin that
will become worse as $N$ increases.  To fix ideas, think of the
quotient of the direct functions: at $x=1/2$,
$g(1/2)/b(1/2)=2^Ne^{-N/2}\approx e^{0.2N}$.


 Eqs.~(\ref{eq:Hii'}) are not fits but checked conjectures that should
 lead to the following CAT (Computer Assisted Theorem):

CAT. {\em Hamiltonian matrices with no hidden symmetries have, except
  near the origin and within fluctuations, canonical Lanczos forms
  that depend only on their dimensionality, second and third moments
  and rank of the interaction}.

No hidden symmetries imply that the matrices cannot be split in
blocks. Which means that eigenvalues cannot be degenerate~\cite[page
300]{wil65}. In other terms: the spectrum must have level repulsion.
If this condition is not fulfilled, degeneracies (not detected by the
Lanczos method), or semi-degeneracies (leading to strong fluctuations)
will make it impossible to define a canonical form. Therefore, it is
here that the condition of good quantum numbers comes in. 

The rank dependence is crucial; e. g., for GOE matrices the equivalent
of Eqs.~(\ref{eq:Hii'}) is $H_{i\,i}=0$ and $H_{i\,i+1}=\sqrt{1-i/d}$
({\em sic}). (The cherry on the CAT cake)

\smallskip

The behaviour of the matrix at the origin depends on the pivot. The
only general statement is that $H_{i\, i+1}$ must be ($O\equiv$ of order)
$O(\sigma)$ at the origin and raise to reach the nib line after $O(N)$
iterations. If the first diagonal is taken to be the lowest
unperturbed state, or chosen variationally, it will be at an energy
$O(N)$ below the log line and raise to reach it after $O(N)$
iterations. It seems a rather general fact that for such ``good''
pivots, diagonals and off-diagonals reach their respective lines in
$\approx N/2$ iterations~\cite{paris,wah99}. It goes without saying
that the correct descriptions of low-lying states and strength
functions hinges on a good description at the origin. The contribution
of Eqs.~(\ref{eq:Hii'}) is nonetheless essential: truncated
calculations will have smaller $\sigma$ and the nib line will be
lower. As a consequence, the ground state energy will not converge to
the correct value and the density of states that determine the final
aspect of the strength functions will be badly missed~\cite{wah99}.

Eqs.~(\ref{eq:Hii'})---eventually supplemented by the rapid raise at the
origin---define an integrable problem, leading to a spectrum of
locally equidistant levels (picket-fence) and locally uniform strength
distributions. When fluctuations are switched on, the level spacings
can be expected to follow a Wigner law, and the strengths a
Porter-Thomas distribution~\cite{bro81}. To put it in other words: the
system becomes chaotic and the strength is localized.

Eqs.~(\ref{eq:Hii'}), duly ``dressed'', amount to a general but
constrained Anderson model, in which constants are replaced by log and
nib functions, while behaviour at the origin and fluctuations (the
dress) must be characterized in terms of properties of the
Hamiltonian. It is likely that the large oscillations, $O(N)$ steps
near the origin---as in Fig.~\ref{fig:sc48}---depend both on the pivot
and on ``details'' of the Hamiltonian. Then, secular trends take over:
fluctuations in the diagonals are about double that of the
off-diagonals, and both seem to follow a random walk in which---at
each step---it is more probable to keep the same direction than to
reverse it.

Perfect dressing amounts to exact solutions, but model dressing is
also possible and can be quite instructive.

For level densities, fluctuations matter less and it is obvious that
the digonalization of the matrices defined by Eqs.~(\ref{eq:Hii'})
will yield densities very close to the exact ones.  It is perhaps not
so obvious that CAT has a corollary:

CAT corollary {\em Level densities at fixed quantum numbers are
binomial.}

\begin{figure}[h]
\begin{center}
  \leavevmode
  \epsfig{file=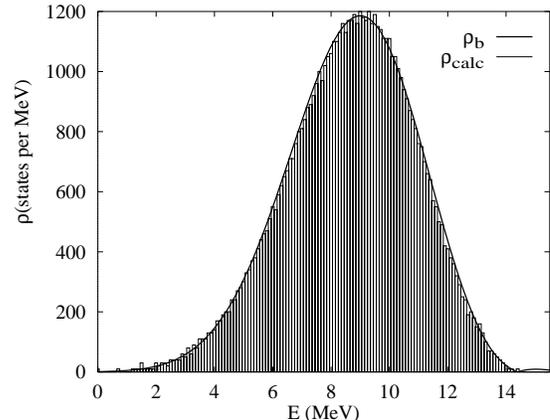, width=8cm}
\caption {Exact level densities for $(fpgd)^8,\, J=6$ compared with
the binomial ones. Parameters from~Fig.~\ref{fig:niba}, treated as
explained in~\protect\cite[Eq. 5]{zuk99} ($d_0=1$)}  
\label{fig:fpgd6.r}
\end{center}
\end{figure}

The checks are fully convincing for all the calculated cases , except
when the dimensionalities become very small ($d \approx 100$).
Fig.~\ref{fig:fpgd6.r} is an example. 
The only significant problem is that the position of the ground
state---which depends critically on the behaviour of the matrix at the
origin---is not predicted with sufficient accuracy for
spectroscopic purposes. 

It should be noted that in Fig.~\ref{fig:fpgd6.r} an Edgeworth
corrected Gaussian\cite{haq90} works as well as the
binomial~\cite{kot00}.  As explained after Eq.~(\ref{eq:G}), binomials
and Gaussians---and {\em a fortiori} Edgeworth corrected ones---can be
close for the small $N$ accessible to simulations. But not in general.
 \begin{figure}[h]
\begin{center}
  \leavevmode
  \epsfig{file=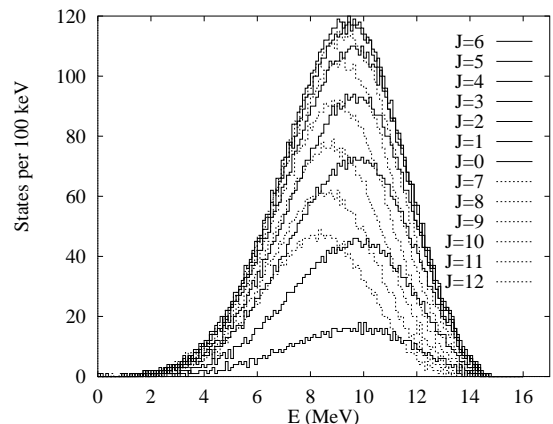, width=8cm}
\caption { Level densities for $(fpgd)^8$,
  $J=0,12$. The legend indicates the order in which $\rho_J$ appear as
  $d_J$ increases (full lines) or decreases (dashes)}
\label{fig:fpgdJ.r}
\end{center}
\end{figure}
For $(fpgd)^8$, the calculated multimodal total density,
$\rho_m=d^{-1}\sum (2J+1)\rho_J$, turns out to be as well described by
a single binomial as the unimodal ones. The gentle buildup shown in
Fig.~\ref{fig:fpgdJ.r} helps in understanding what happens. For
$^{48}$Ca the situation is the same. We have two examples of a
multimodal $\rho_m$ behaving as unimodal, which squares with the
systematic success of Bethe's formula~\cite{zuk99}.  Assuming that the
signal for an eventual phase transitions is a significant deviation
from binomial behaviour, the conclusion is that no low energy thermal
phase transitions occur in nuclei.

\begin{figure}[htb]
\begin{center}
  \leavevmode
  \epsfig{file=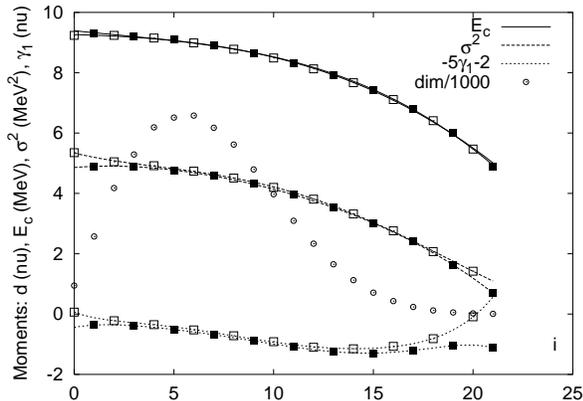, width=8cm}
\caption { Centroid $E_c$,
  $\sigma^2$, $-5\gamma_1-2$ and $d/1000$ for $(fpgd)^8$. Full (open)
  squares for $J$ odd (even).}
\label{fig:fpgdJm}
\end{center}
\end{figure}

For the individual $\rho_J$, Bethe's expression~\cite[Eq.(2B-62)]{BM69}
amounts to $E_c\equiv J(J+1)$ and constant $\sigma^2$, while---for
$(fpgd)^8$---in Fig.~\ref{fig:fpgdJm}, we find that both $E_c$ and
$\sigma^2$ are nearly parabolic, the former with opposite sign to
Bethe's. For $^{48}$Ca, $E_c$ is nearly constant and $\sigma^2$ again
an inverted parabola. No reason for alarm, but the subject demands
further study.
One thing is clear: we have to learn how to calculate low moments at
fixed quantum numbers. A brute force approach is viable in very large
spaces, but a bit of formalism could extend the range to arbitrarily
large ones. The dimensionalities in Fig.~\ref{fig:fpgdJm} provide an
example: they are the differences of two binomials, obtained by
realizing that the calculation of the number of states at fixed $J_z$
is a single particle problem.

Once we have the moments, many things can be done. One of them is to
treat random Hamiltonians and understand the canny insistence of $J=0$
to be ground state~\cite{JO99}. Probably the most promising is the
calculation of thermodynamic quantities.   

Thermodynamically, the behaviour of a single binomial is elementary.
Write $\rho(n)=\lambda^n{N \choose n}$, with $E=n\epsilon$.  Then
$z(\beta)=\sum_n \rho(n)e^{-\beta \varepsilon n}=(1+\lambda e^{-\beta
  \varepsilon})^N,$ is the partition function and $E(\beta)=\partial
\ln z/\partial \beta=N\varepsilon \lambda \, e^{-\beta
  \varepsilon}/(1+\lambda\, e^{-\beta \varepsilon})$, the thermal
energy: The same as obtained from a microcanonical calculation,
$\beta=\partial \ln \rho/\partial E$, using the Stirling approximation
for $\rho$.  Therefore, continuous and discrete binomials have the
same thermodynamics; The canonical partition function for a general
problem is enormously simplified, as it becomes the sum over $x$, a
set of conserved quantum numbers ($E_{0x}$ is the energy origin):
\begin{equation}
    \label{eq:Z}
    Z(\beta)=\sum_x\left (1+\lambda_x\,e^{-\beta\varepsilon_x}\right
    )^{N_x}e^{\beta E_{0x}},  
\end{equation}
More on this soon~\cite{dzxx}.

\vspace{.1cm}

Let us close by making explicit the striking complementarity between
the approach in the previous paper, involving scalar traces
$d^{-1}\langle {\cal H}^{K} \rangle$, and the Lanczos approach here,
involving expectation values for a pivot at fixed quantum numbers,
$\langle 0|{\cal H}^{K}|0\rangle$. Low values of $K$ describe the
centroid region, and are easy to find for the traces, while they
describe the ground state in Lanczos and are hard to calculate. The
converse is---or at least seems to be---true for high $K$. CAT is more
general and has far more consequences than the CLT
results~\cite{zuk99}, but they are theorems, while CAT is a
conjecture. The theorems will be of help if they can be extended to
traces at fixed quantum numbers. A task we have to tackle at any rate.

\end{multicols}
\end{document}